\documentclass[aps,pra,twocolumn,showpacs,superscriptaddress]{revtex4-1}
\usepackage{amsmath,amssymb,graphicx}
\usepackage{float}
\begin{document}

\title{Grand-canonical phase diagram and chiral-current suppression at $\pi$ flux in a bosonic two-leg ladder} 

\author{Meng Zhang}
\affiliation{School of Physics,  Anhui University, Hefei 230601, China}
\author{Qingyun Xu}
\affiliation{School of Physics,  Anhui University, Hefei 230601, China}
\author{Zhi Lin}
\email{zhilin18@ahu.edu.cn}
\affiliation{School of Physics, Anhui University, Hefei 230601, China}
\affiliation{State Key Laboratory of Surface Physics, Fudan University, Shanghai 200433, China}

%\date{\today}

\begin{abstract}
We investigate the ground-state phase diagram of repulsively interacting bosons on a two-leg ladder threaded by a uniform artificial magnetic flux, using the cluster Gutzwiller mean-field method. In the strong-rung-coupling regime, self-consistent calculations are performed on a $2\times4$ cluster. By analyzing the superfluid order parameter, leg-resolved currents, chiral current, the ratio of currents on adjacent legs, and the density imbalance between the two legs, we distinguish Mott-insulating regions from superfluid regimes and further characterize the observed states as Meissner-like, vortex-like (superfluid or Mott insulating), or biased-ladder states across a broad range of  parameters. In the parameter regions accessible to previous density-matrix renormalization-group studies, our results show qualitative agreement with the established phase structure, demonstrating that the cluster Gutzwiller approach provides a useful compromise between computational efficiency and physical accuracy. Building on this benchmark, we construct, to our knowledge, the first grand-canonical $t$--$\mu$ phase diagrams for this flux ladder system. These diagrams reveal how the magnetic flux modifies the shape, tilt, and extent of the Mott lobes. We further explore parameter regimes that have remained largely inaccessible in previous studies, including higher fillings $\rho\gtrsim1$ and the intermediate interaction window $U/t\in[7.69,9.09]$. Particular attention is paid to the special flux point $\varphi=\pi$, where the flux is equivalent to $-\pi$ modulo $2\pi$ and the effective triangular-ladder mapping becomes singular. Within our cluster Gutzwiller solutions, this symmetry constraint is reflected in the absence of net chiral currents and in the stabilization of a nonchiral Mott-insulating state in the relevant parameter regime, in contrast to the chiral-superfluid tendency expected away from $\varphi=\pi$. Our results provide a computationally efficient route for mapping the global phase structure of bosonic flux ladders and offer guidance for future ultracold-atom experiments in artificial gauge fields.

\end{abstract}

\maketitle

\section{Introduction}
Optical lattices  loaded with ultracold atoms provide a highly controllable platform  for quantum simulation of strongly correlated many-body systems~\cite{Lewenstein2012,Bloch2008}. The lattice geometry, dimensionality, tunneling amplitudes, and interaction strengths can be engineered with a high degree of flexibility, while interatomic interactions may be tuned via  Feshbach resonances~\cite{Feshbach}, and Floquet engineering~\cite{Floquet0,Floquet1,Floquet2,Floquet3}. Since the prediction and experimental observation of the superfluid--Mott-insulator transition in the Bose-Hubbard model~\cite{Jaksch1998,Greiner2002}, optical lattices have become a central arena for probing correlated bosonic phases in a variety of lattice geometries~\cite{Windpassinge,zhi-3,zhi-4,Yue} and interaction regimes~\cite{non_standard,non_standard1,zhi-5}. Building on these advances, rapid progress in realizing artificial gauge fields for neutral atoms, both in discrete lattice systems~\cite{Yasunaga2007,Williams2010,Aidelsburger2013,Miyake2013} and in continuum settings~\cite{Compton2009,Jimenez2009}, has opened a route to studying orbital magnetism and flux-induced quantum phase transitions in highly controllable many-body systems.

The ladder systems constitute the minimal lattice geometry capable of exhibiting  the effects of artificial gauge fields on ultracold atoms~\cite{Halati2023}. As quasi-one-dimensional structures intermediate between purely one- and  two-dimensional systems, ladders occupy a distinctive position in quantum many-body physics, offering a favorable setting for investigating the interplay among interactions, geometry, and gauge fields~\cite{Halati2023,Dagotto1996,Cazalilla2011}. Among them, the two-leg ladder is of  particular importance because it is the simplest geometry that supports plaquette flux and leg-resolved currents, while still exhibiting physics that is qualitatively distinct from a single chain~\cite{Halati2023,Donohue2001}. When an artificial magnetic field is  introduced, the flux is incorporated into the Bose-Hubbard model via complex tunneling couplings~\cite{Oktel2007,Umucalilar2007,Goldbaum2008,Sachdeva2010,Sachdeva2012,Powell2011},  giving rise to a variety of novel quantum phases, including the Meissner phase, vortex phases, vortex-lattice states, and the biased-ladder 
phase~\cite{Sachdeva2017,Deng2015,Tokuno2014,Strinati2018,Orignac2016,Piraud2015,Haller2020,Sachdeva2018,Buser2020,Uchino2016,Ceven2022,Natu2015,Orignac2001,Keles2015,Atala2014,Mancini2015}. The emergence of these phases reflects characteristic orbital responses, most notably chiral edge currents and vortex structures: in the Meissner regime the flux is effectively expelled from the bulk, whereas in the vortex regime it penetrates the ladder in the form of quantized circulation patterns~\cite{Dio2015}. Their stability is further shaped by the interplay among orbital frustration, interactions, and the ladder geometry.

Theoretical studies of interacting bosons on flux ladders have so far predominantly relied on the  density matrix renormalization group (DMRG) method~\cite{Dhar2013,Dhar2012,Greschner2016,Halati2023,Barbiero2023}, which provide highly accurate results in one-dimensional and quasi-one-dimensional geometries. Among them, Greschner et~al.
performed a detailed study of this model at intermediate interaction strengths ($U/t \in (1,10)$), revealing stable vortex lattice states (with vortex densities $\rho_v = 1/2, 1/3, 1/4$), a biased-ladder superfluid phase, and a charge-density-wave phase, and predicting a chiral-current reversal effect associated with the vortex lattices~\cite{Greschner2016}. However, most existing numerical studies have focused on fixed-density calculations 
and on relatively low filling factors, typically $\rho\leq1$. Extending DMRG calculations 
to higher fillings can become computationally demanding because of the enlarged local 
Hilbert space, increased entanglement, and the difficulty in converging metastable states. 
As a result, the high-filling regime and the global phase structure in the grand-canonical 
$t$--$\mu$ plane remain much less explored.

A complementary perspective was recently proposed by Barbiero et al.~\cite{Barbiero2023}, who showed that in the strong-rung-coupling regime and for flux close to $\pi$, a two-leg square ladder can be mapped onto an effective triangular ladder with staggered flux. The resulting low-energy physics is described by a frustrated spin-$1/2$ quantum XX model, 
providing an appealing interpretation of the vortex-lattice insulating phase (corresponding to a bond-ordered-wave phase) and the biased-ladder superfluid phase (corresponding to a chiral superfluid phase) in terms of frustrated magnetism. This mapping, however, becomes singular at the special point $\varphi=\pi$, where the effective nearest-neighbor tunneling amplitude $t_1 \propto -t\cos(\varphi/2)$ vanishes. The physics exactly at $\varphi=\pi$ therefore requires a direct treatment of the original square-ladder model rather than an extrapolation from the effective triangular-ladder description. Since $\varphi=\pi$ is equivalent to $-\pi$ modulo $2\pi$, this point also possesses an additional symmetry constraint on the chirality of the system, forbidding  the formation of chiral currents. Clarifying the nature of the ground state at this point is thus important for completing the phase diagram near maximal flux frustration.

These considerations motivate the use of complementary numerical methods that can efficiently access broad parameter regimes while retaining essential short-range quantum correlations. The cluster Gutzwiller mean-field (CGMF)~\cite{Gutzwiller0,Gutzwiller1,Gutzwiller2,hexagonal-Gutzwiller,Gutzwiller3,Gutzwiller4,Lin2026} method provides such an approach by treating quantum correlations exactly within a finite cluster and incorporating intercluster couplings at the mean-field level. Compared with a single-site Gutzwiller approximation, it captures local correlation effects and current patterns within the cluster; compared with DMRG, it is computationally less demanding for broad parameter scans and is naturally suited to grand-canonical calculations. These features make it particularly useful for mapping global phase diagrams at high fillings and extreme flux values. CGMF method has been successfully applied to Bose-Hubbard ladder systems without magnetic flux~\cite{Sachdeva2017,Deng2015,Sachdeva2018}. 
More recently, Lin et al.~\cite{Lin2026} combined cluster Gutzwiller calculations with a machine-learning-based $\Delta$-learning strategy, demonstrating an efficient route to high-resolution phase diagrams in complex Bose-Hubbard systems.

In this work, we apply the  CGMF  method to interacting bosons on a two-leg ladder threaded by a uniform magnetic flux. We focus on the strong-rung-coupling regime and perform self-consistent calculations on a $2\times4$ cluster. Several physical observables are used to characterize the phases, including the superfluid order parameter, leg-resolved particle currents, current ratios on adjacent legs, and the density imbalance between the two legs. In parameter regions overlapping with previous DMRG studies, our results reproduce the known qualitative phase structure, thereby benchmarking the cluster Gutzwiller approach for this problem.

Our work makes three main contributions. First, we construct, to our knowledge, the first grand-canonical $t$--$\mu$ phase diagrams for this bosonic flux ladder system, providing a global perspective that is difficult to obtain directly with DMRG. These diagrams show how the artificial magnetic flux modifies the shape, tilt, and extent of the Mott lobes. Second, we extend the exploration to parameter regimes that have remained largely unexplored, including higher fillings $\rho\gtrsim1$ and the intermediate interaction window $U/t\in[7.69,9.09]$. Third, we analyze the special flux point $\varphi=\pi$, where the effective triangular-ladder mapping breaks down. Within our cluster Gutzwiller solutions, the symmetry equivalence between $\varphi=\pi$ and $\varphi=-\pi$ is manifested by the disappearance of net chiral currents and by the emergence of a nonchiral Mott-insulating state in the relevant parameter regime. This behavior contrasts with the chiral-superfluid tendency expected away from $\varphi=\pi$ and highlights the singular nature of the exactly $\pi$-flux point.

Together, these results establish the cluster Gutzwiller approach as a useful tool for mapping the global phase structure of bosonic flux ladders over broad parameter regimes. The remainder of this paper is organized as follows. In Sec.~\ref{Model}, we introduce the two-leg semisynthetic bosonic flux ladder model, outline the CGMF formalism, and define the key observables used to characterize the ground-state phases. In Sec.~\ref{Results}, we present the numerical phase diagrams and analyze the corresponding order parameters, currents, density imbalance, and special behavior at $\varphi=\pi$. Finally, Sec.~\ref{Summary} summarizes our results and discusses possible extensions.
\begin{figure}[htbp]
		\centering
		\includegraphics[width=0.90\linewidth]{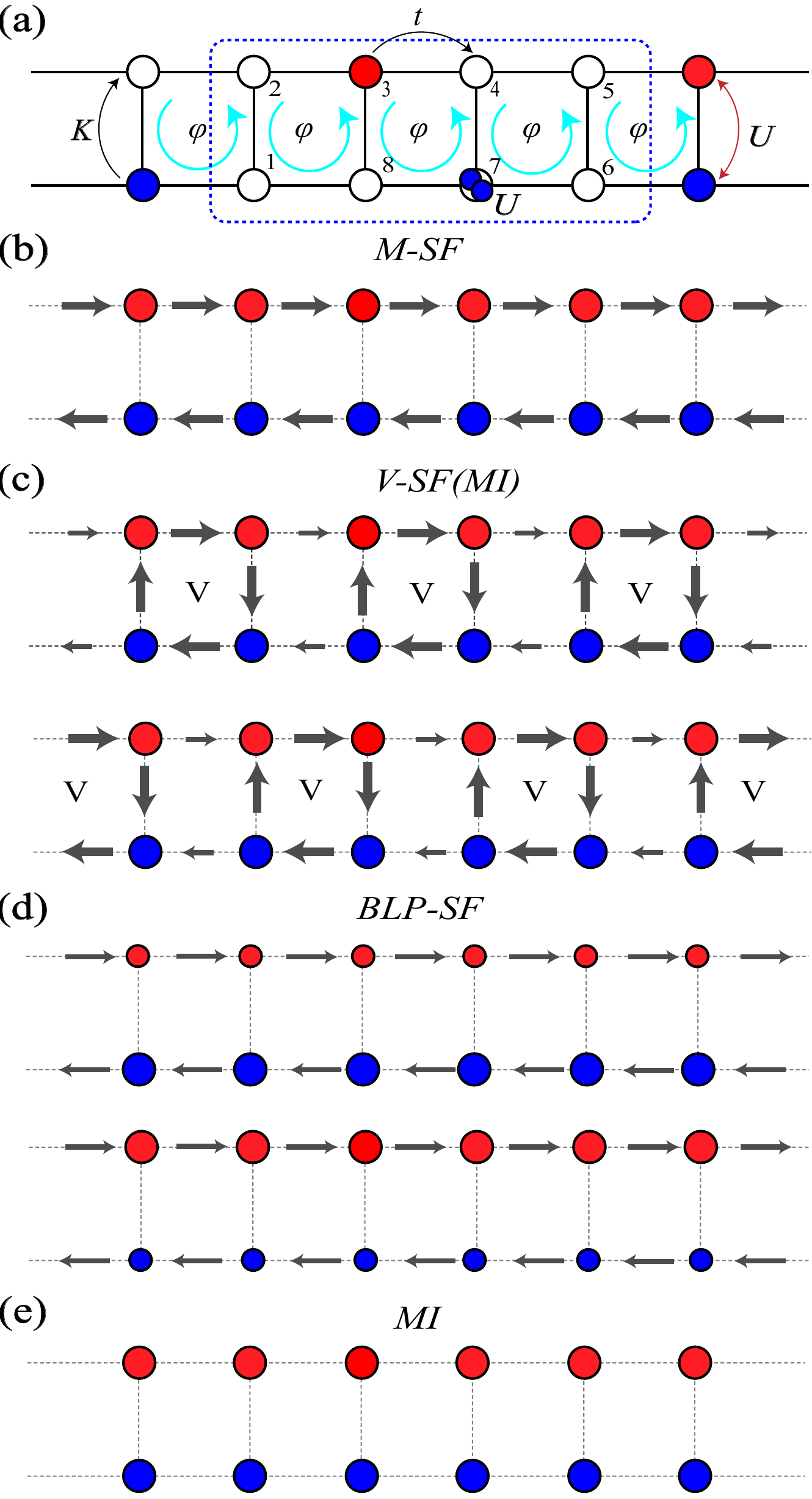}
		\caption{(a) Sketch of the two-leg semisynthetic flux ladder, showing intra-leg and inter-leg tunnelings $t$ and $K$, equal on-site and nearest-neighbor interactions $U_{\sigma,\sigma}=U_{\sigma,-\sigma}$, and the magnetic flux $\varphi$.(b)-(e) Sketch of current (arrows) and density (circles) distributions for different quantum phases: (b) Meissner superfluid (M-SF), (c) two typical vortex configurations (each of which can be either superfluid or Mott insulating depending on the parameter regime), (d) biased-ladder superfluid (BLP-SF), (e) nonchiral Mott insulator (MI).}
		\label{fig:zm}
	\end{figure}

\section{Model, Method, and observables}\label{Model}
As shown in Fig.~\ref{fig:zm}, we consider a bosonic ladder consisting of one real spatial dimension and one  synthetic spin dimension. The synthetic dimension is formed by two internal  atomic states $\sigma = \pm 1/2$ coupled via Raman transitions, which serve as 
the second leg of the ladder. A uniform artificial magnetic flux is threaded  through the ladder, realizing a so-called semisynthetic flux  ladder~\cite{Celi2014,Greschner2016,Barbiero2023}. The corresponding  Hamiltonian reads
	\begin{equation}
		\begin{aligned}
			H & =-t\sum_{j,\sigma } \left ( e^{-i\varphi \sigma  } a_{ j+1,\sigma } ^{\dagger}a_{ j,\sigma } +H.c. \right )     \\
			& \quad +K\sum_{j}\left ( a_{ j,\sigma } ^{\dagger}a_{ j,-\sigma } +H.c. \right ) +\frac{U_{\sigma,\sigma }}{2}\sum_{j,\sigma } \! n_{ j,\sigma }\left ( n_{ j,\sigma }\!-\!1  \right )  \\
		    & \quad +\frac{U_{\sigma,-\sigma }}{2}\sum_{j } n_{ j,\sigma } n_{ j,-\sigma}  +\mu \sum_{j,\sigma} n_{ j,\sigma }.	  
		\end{aligned} 
		\label{eq:bose-Hubbard}
	\end{equation}
	Here $a_{j}$ is the bosonic annihilation operator at lattice site $j$. The intra-leg tunneling is described by a hopping amplitude $t$ with a spin-dependent Peierls phase $\phi_\sigma = \varphi \sigma$, where $\varphi$ represents the classical magnetic flux per plaquette acquired from the Raman momentum transfer, while the inter-leg tunneling is denoted by $K$.  The intra-leg on-site interaction $U_{\sigma,\sigma}$ and the inter-leg 
nearest-neighbor interaction $U_{\sigma,-\sigma}$ are set equal to each  other and taken as unity in this work.

To accurately characterize the ground-state properties of this quantum many-body system, we employ the CGMF method, which has been widely used in studies of various nonstandard Bose-Hubbard models. Since this approach has been extensively documented for interacting bosonic systems, we refrain from repeating the 
algorithmic details here (see, e.g., Refs.~\cite{Gutzwiller2,Gutzwiller4,Lin2026} for comprehensive descriptions) and only outline the central idea: by exactly diagonalizing a finite-size cluster and self-consistently solving the mean-field boundary conditions, the method captures short-range spatial correlations and achieves an accuracy significantly beyond that of the conventional 
single-site mean-field approximation.

First, we compute the superfluid order parameter $\langle a \rangle$, which signals the breaking of the $U(1)$ gauge symmetry, to distinguish the Mott-insulating (MI) phase from the superfluid (SF) phase. In the MI phase $\langle a \rangle = 0$, while in the SF phase $\langle a \rangle \neq 0$. The particle filling $\rho$ is also evaluated in both phases. Next, we examine a class of important quantities that characterize the current patterns induced by the artificial magnetic flux. The local currents are described by the leg current $J_{\parallel}$ and the rung current $J_{\perp}$, defined as
\begin{equation}
\begin{gathered}
J_{j,\sigma}^{\parallel} = -it\left( e^{-i\varphi\sigma}a_{j+1,\sigma}^{\dagger}a_{j,\sigma} 
- e^{i\varphi\sigma}a_{j,\sigma}^{\dagger}a_{j+1,\sigma} \right), \\
J_{j}^{\perp} = ik\left( a_{j,\sigma}^{\dagger}a_{j,-\sigma} - a_{j,-\sigma}^{\dagger}a_{j,\sigma} \right).
\end{gathered}
\end{equation}
To further resolve the vortex structure, we introduce the ratio of currents on adjacent 
legs $I$, given by
\begin{equation}
I = \frac{ e^{-i\varphi/2} a_{2}^{\dagger}a_{3} - e^{i\varphi/2} a_{3}^{\dagger}a_{2} }
{ e^{-i\varphi/2} a_{3}^{\dagger}a_{4} - e^{i\varphi/2} a_{4}^{\dagger}a_{3} }.
\end{equation}
In the Meissner phase, the currents on adjacent legs are uniform and parallel, leading 
to $I = 1$; by contrast, the formation of a vortex phase introduces an alternating 
current pattern, which necessarily drives $I$ away from unity.  A clear deviation of 
$I$ from $1$ therefore signals the onset of a vortex state.

In addition, we compute the nonlocal chiral current $J_{c}$, which quantifies the net 
chiral asymmetry of the system and is defined as~\cite{Greschner2016}
\begin{equation}
J_{c} = \frac{1}{N} \sum_{j} \left| J_{j,\sigma}^{\parallel} - J_{j,-\sigma}^{\parallel} \right|.
\end{equation}
Finally, we evaluate the density imbalance between the two legs. For the $2\times 4$
cluster adopted in this work (corresponding to the blue dashed region in Fig.~1(a)),
this quantity is defined as
\begin{equation}
\Delta n = n_{2} + n_{3} + n_{4} + n_{5} - n_{1} - n_{8} - n_{7} - n_{6},
\end{equation}
which captures the breaking of the $Z_{2}$ symmetry between the legs and is used to
identify biased-ladder phases.

Based on the definitions of the key observables introduced above, we summarize 
the characteristic properties of the quantum phases appearing in this work in 
Table~\ref{tab:quantum_phases}.

 	\begin{table}[htbp]
 		\centering
 		\renewcommand{\arraystretch}{1.3}   % increase row height
\caption{Characteristic properties of the identified quantum phases in terms of the 
key observables defined in the text.}
 		\label{tab:quantum_phases}
 		\begin{tabular*}{0.42\textwidth}{@{\extracolsep{\fill}} c c c c c@{}}
 			\hline
 			\textbf{Quantum phase} & $\langle a \rangle$ & $J_{\perp}$ & $I$ & $\Delta N$ \\
 			\hline
            M-SF   & $\neq 0$                  & $10^{-5} \sim 10^{-4}$ & 1   & 0   \\
           % M-MI   & 0                         & $10^{-5} \sim 10^{-4}$ & 1   & 0   \\
            V-SF   & $\neq 0$                  & $10^{-3}$              & $\neq 1$ & 0   \\
            V-MI   & 0                         & $10^{-3}$              & $\neq 1$ & 0   \\
            BLP-SF & $\neq 0$                  & $10^{-5} \sim 10^{-4}$ & 1   & $\neq 0$ \\
 			\hline
 		\end{tabular*}
 	\end{table}

\section{Numerical Results}\label{Results}
		\begin{figure}[htbp]
	\centering
	\includegraphics[width=0.99\linewidth]{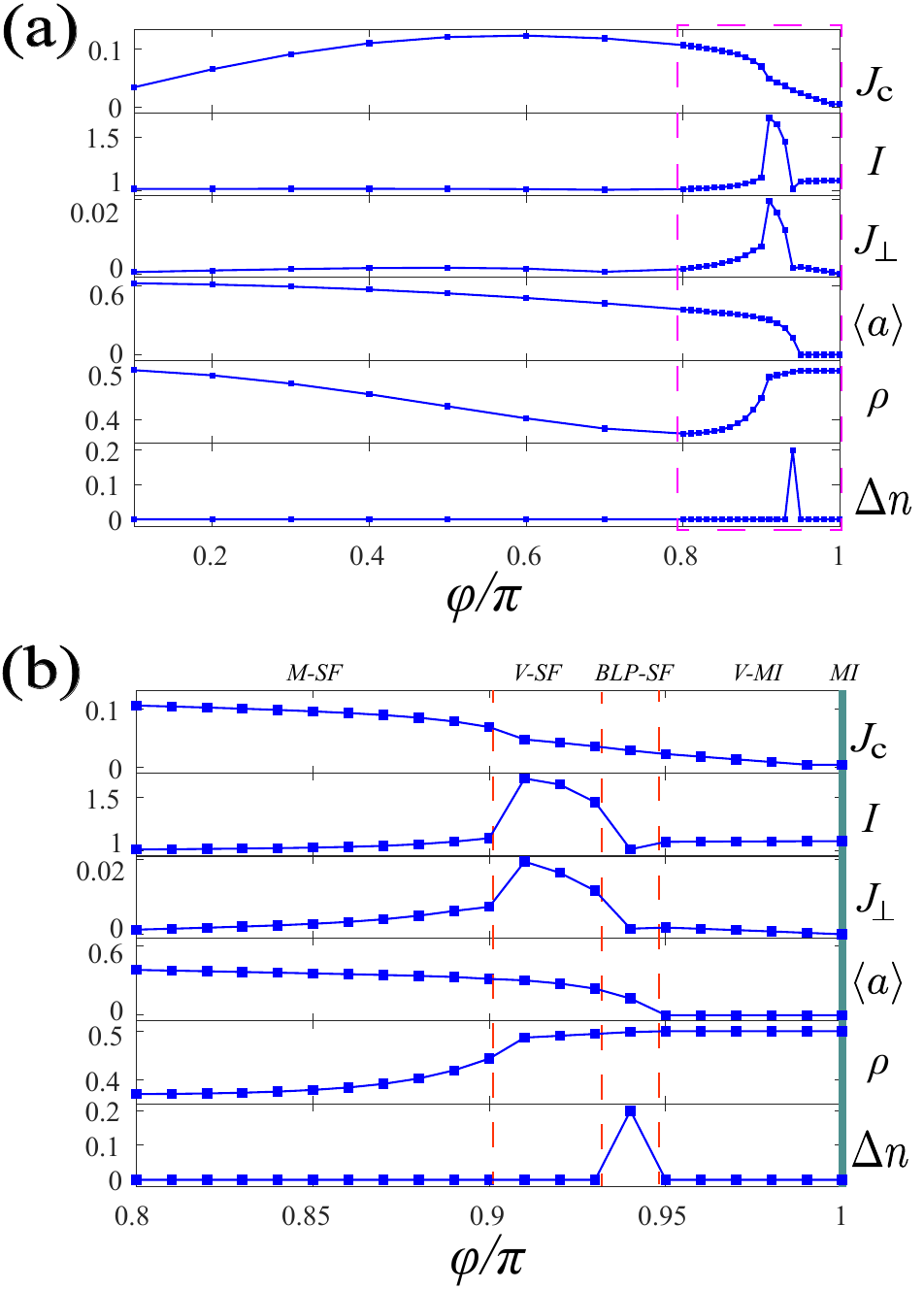}
	\caption{(a) Key observables $J_{c}$, $J_{\perp}$, $I$, $\langle a\rangle$, $\rho$, and  $\Delta n$ as functions of $\varphi/\pi$ obtained from CGMF calculations at  $K=10t$, $\mu=-2.47$, and $t=0.252$. (b) Enlarged view of the region enclosed by the magenta dashed box in (a).}
	\label{fig:xu}
\end{figure}
To benchmark the effectiveness of the CGMF method for studying bosonic flux ladders, we perform numerical simulations on a $2\times 4$ cluster in the strong-rung-coupling regime ($K=10t$). We choose a parameter region close to that investigated by Barbiero et al.\ using DMRG~\cite{Barbiero2023}: whereas their study focused on fixed filling $\rho=1/4$ and intermediate interactions $U/t\in[0,4]$, we work in a low-filling regime $\rho\in[0.33,0.5]$ with intermediate interaction $U/t\approx3.968$ and chemical potential $\mu=-2.47$. For these parameters, we calculate several key observables as functions of the magnetic flux $\varphi$, with 
the results shown in Fig.~\ref{fig:xu}.

From Fig.~\ref{fig:xu}, we find that as the flux $\varphi$ increases, the ground state evolves  successively from the Meissner superfluid (M-SF) phase, to the vortex superfluid  (V-SF) phase, to the biased-ladder superfluid (BLP-SF) phase, to the vortex Mott  insulator (V-MI) phase, and finally to a Mott insulator (MI) at $\varphi=\pi$.  The identified quantum phases are in qualitative agreement with the DMRG results of Barbiero et al.~\cite{Barbiero2023}. We note that, due to the limited cluster size accessible in CGMF calculations, the present method cannot resolve the incommensurate-to-commensurate transition between the vortex liquid and the vortex lattice. Therefore, in this work we do not distinguish between these two vortex states and refer to them collectively as vortex phases (either V-SF or V-MI). The qualitative consistency with DMRG in the overlapping parameter region validates the CGMF method as a reliable tool for studying this system.
\begin{figure}[htbp]
	\centering
	\includegraphics[width=0.99\linewidth]{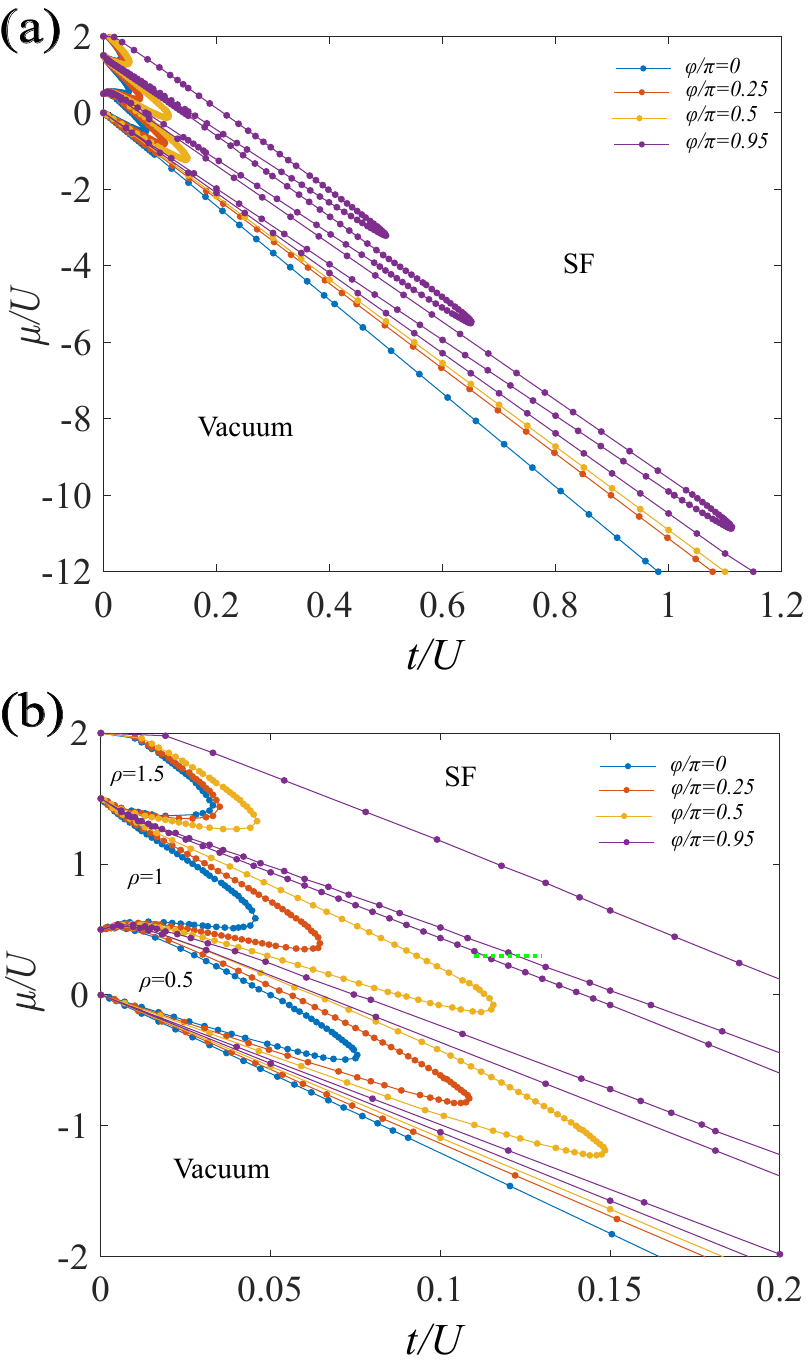}
	\caption{(a)The grand-canonical $t$--$\mu$ phase diagrams obtained from CGMF calculations  at $K=10t$ for $\varphi/\pi=0$ (blue), $0.25$ (red), $0.5$ (yellow), and  $0.95$ (purple). Data points are connected by lines as a guide to the eye. (b) Enlarged view of (a).}
	\label{fig:TMU}
\end{figure}

Having validated the CGMF method, we now turn to parameter regimes that lie beyond  the reach of previous DMRG studies by Barbiero et al., namely the high-filling  regime $\rho\gtrsim1$ and a wider intermediate interaction window  $U/t\in[7.69,9.09]$. We first compute the grand-canonical $t$--$\mu$ phase diagrams at $K=10t$ for $\varphi/\pi=0,\,0.25,\,0.5,\,0.95$, which, to our knowledge, are 
the first such phase diagrams for this bosonic flux ladder system and offer a global  view that is challenging to obtain directly with DMRG. The results are displayed in Fig.~\ref{fig:TMU}. We observe a clear modification of the MI--SF phase boundary with increasing flux: the MI region expands continuously, and, in contrast to the zero-flux case, the Mott lobes become significantly tilted as the flux grows.

\begin{figure}[H]
	\centering
	\includegraphics[width=0.99\linewidth]{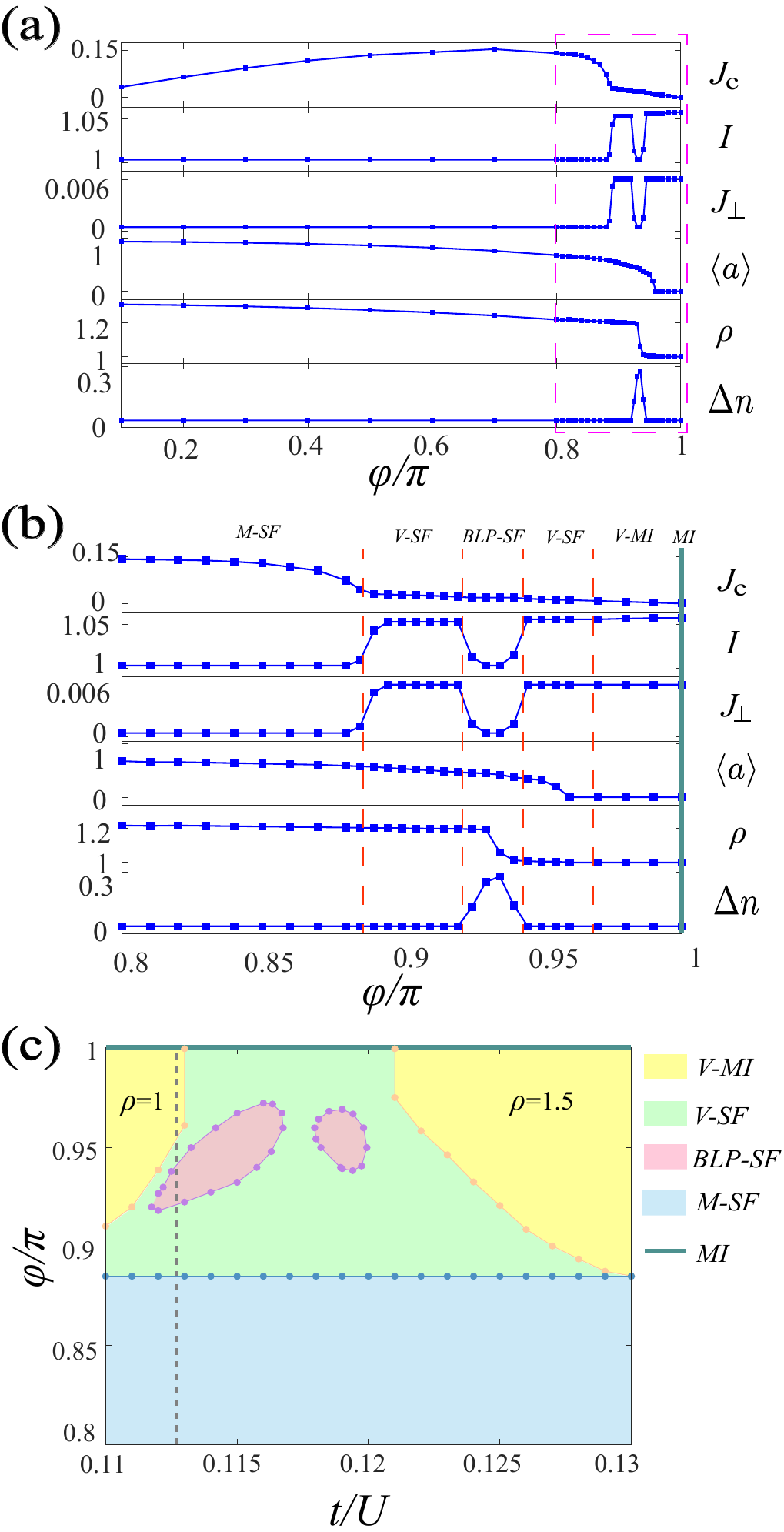}
	\caption{Phase diagram.(a) Key observables $J_{c}$, $J_{\perp}$, $I$, $\langle a\rangle$, $\rho$, and $\Delta n$ 
as functions of $\varphi/\pi$ in the high-filling regime, obtained from CGMF calculations at $K=10t$, $\mu=0.3$, and $t=0.11275$.  (b) Enlarged view of the region enclosed by the magenta dashed box in (a).  (c) Phase diagram of the two-leg semisynthetic flux ladder in the $t$--$\varphi$ plane in the high-filling regime, obtained from the CGMF method for $K=10t$ and $\mu=0.3$. The gray dashed line at $t = 0.11275$ indicates the fixed-$t$ cut corresponding to the parameter points of the curves in (a).}
	\label{fig:tn}
\end{figure}

Subsequently, we further compute the $t$--$\varphi$ phase diagram in the high-filling regime ($\rho\gtrsim1$) for $t\in[0.11,0.13]$ (corresponding to $U/t\in[7.69,9.09]$, i.e., the green line segment in Fig.~3 (b) with $\mu=0.3$) and $\varphi/\pi\in[0.8,1]$. As shown in Fig.~\ref{fig:tn}(c), for $\varphi<0.88\pi$ the system resides in the M-SF phase.
As $\varphi$ increases further, it enters the V-SF phase, and at sufficiently large $\varphi$, the V-MI phase appears. Notably, islands of the BLP-SF phase emerge within the V-SF region; to our knowledge, this phenomenon is revealed for the first time. The BLP-SF phase is characterized by $\Delta n\neq0$, indicating an imbalanced particle distribution between the two legs and the breaking of the $Z_{2}$ symmetry between them.

Our $t$--$\varphi$ results exhibit slight differences compared to the DMRG findings of Barbiero et al.~\cite{Barbiero2023}. These differences arise because their DMRG calculations are performed in the canonical ensemble at a fixed filling $\rho=1/4$ and at intermediate interactions $U/t\in[0,4]$, whereas the CGMF method operates in the grand-canonical ensemble with a variable total particle number and accesses higher fillings $\rho\gtrsim1$ and a higher interaction window $U/t\in[7.69,9.09]$. Along the gray dashed line in Fig.~\ref{fig:tn}(c) ($t=0.11275$), we calculate $J_{c}$, $J_{\perp}$, $I$, $\langle a\rangle$, $\rho$, and $\Delta n$ as functions of $\varphi/\pi$ to discriminate the ground states; these results are displayed in Figs.~\ref{fig:tn}(a) and (b) [where (b) is an enlarged view of the region enclosed by the magenta dashed box in (a)].

At $\varphi/\pi=1$, the effective triangular-ladder mapping becomes singular. At this special flux, the system possesses a joint symmetry under time reversal (denoted by $\mathcal{T}$) and a gauge transformation (denoted by $\mathcal{G}$). Under time reversal, the flux changes sign, $\varphi\to-\varphi$, and the Hamiltonian transforms as $\mathcal{T}H(\varphi)\mathcal{T}^{-1}=H(-\varphi)$, while the chiral current transforms as $\mathcal{T}J_c(\varphi)\mathcal{T}^{-1}=-J_c(-\varphi)$. At $\varphi=\pi$, the Hamiltonians at $+\pi$ and $-\pi$ flux are gauge equivalent; the gauge transformation $\mathcal{G}$ maps the $-\pi$ flux representation back to the $+\pi$ flux representation, satisfying $\mathcal{G}H(-\pi)\mathcal{G}^{-1}=H(\pi)$ and $\mathcal{G}J_c(-\pi)\mathcal{G}^{-1}=J_c(\pi)$. Consequently, the combined antiunitary operation $\mathcal{S}=\mathcal{G}\mathcal{T}$ is a symmetry of $H(\pi)$, and the chiral current is odd under it: $\mathcal{S}J_c(\pi)\mathcal{S}^{-1}=-J_c(\pi)$. For a nondegenerate ground state preserving this symmetry, the expectation value of the chiral current must therefore 
vanish, $\langle J_c\rangle=0$. In our CGMF calculations, no stable symmetry-broken chiral solution is found in the considered parameter regime, leading to a nonchiral MI state at $\varphi=\pi$. This behavior contrasts sharply with the chiral-superfluid tendency expected away from $\varphi=\pi$, and it highlights the singular nature of the exactly $\pi$-flux point.

\section{Summary and outlook}\label{Summary}
We have employed the CGMF method to investigate the ground-state phase diagram of interacting bosons on a two-leg semisynthetic ladder threaded by a uniform artificial magnetic flux.  By performing self-consistent calculations on a $2\times4$ cluster in the strong-rung-coupling regime, we characterized the emergent quantum phases through the superfluid order parameter, leg-resolved and chiral currents, the ratio of currents on adjacent legs, and the density imbalance between the two legs.  Benchmarking against existing DMRG results in the low-filling, intermediate interaction regime confirmed that the CGMF approach captures the essential physics, including the Meissner, vortex, biased-ladder, and Mott-insulating phases, thereby validating its use as an efficient complementary tool for this system.

Using this validated method, we constructed, to our knowledge, the first grand-canonical $t$--$\mu$ phase diagrams for the bosonic flux ladder, which directly reveal how the magnetic flux modifies the Mott lobes—both their extent 
and their tilt.  We further extended the exploration to previously unexplored parameter regimes, including high fillings $\rho\gtrsim1$ and the intermediate interaction window $U/t\in[7.69,9.09]$, and mapped the corresponding 
$t$--$\varphi$ phase diagram.  In doing so, we uncovered island-like biased-ladder superfluid phases embedded within the vortex-superfluid region, a phenomenon not reported before.

A central focus of this work was the special flux point $\varphi=\pi$.  At this point the effective triangular-ladder mapping breaks down, and a combined time-reversal and gauge symmetry ($\mathcal{S}=\mathcal{G}\mathcal{T}$) 
constrains the chiral current to vanish in any nondegenerate symmetric ground state.  Our CGMF solutions explicitly exhibit this behavior: we find a nonchiral MI state at $\varphi=\pi$, in stark contrast to the chiral-superfluid 
tendency that prevails away from this point.  This result clarifies the physical origin of the singularity of the triangular-ladder mapping and completes the phase diagram near maximal flux frustration.

Overall, our study demonstrates that the CGMF method strikes a favorable balance between accuracy and computational efficiency, enabling systematic scans of broad parameter spaces that are challenging for DMRG.  The phase diagrams and the insights gained at $\varphi=\pi$ provide valuable guidance for future ultracold-atom experiments aiming to realize and probe these exotic quantum phases in synthetic gauge fields.  An interesting extension of this work would be to incorporate machine-learning-enhanced CGMF schemes to reach even higher resolution and to address the commensurate–incommensurate vortex transitions that remain beyond the reach of the present cluster sizes.

\section{Acknowledgements}\label{Acknowledgements}
 This research is supported by the National Natural Science Foundation of China (NSFC) under Grant No.12004005, the Scientific Research Fund for Distinguished Young Scholars of the Education Department of Anhui Province No.2022AH020008, and the open project of the state key laboratory of surface physics in Fudan University under Grant No. KF2021$\_$08.

\end{document}